# The Impact Of Social Media In The Fight Against The Spread Of Coronavirus (Covid-19) Pandemic In Anambra State, Nigeria

Okechukwu Christopher Onuegbu[1], Joseph Oluchukwu Wogu[1], Jude Agbo[1*]

Department of Mass Communication, University of Nigeria Nsukka.

**Abstract**

This study examined the impact of social media in the fight against the spread of coronavirus (COVID-19) pandemic in Anambra state, Nigeria. The key objectives are to: find out if the numbers of social media users increased in Anambra state since the wake of coronavirus pandemic; find out if the social media is being utilised in the fight against the spread of coronavirus pandemic in Anambra state; find out how the social media is being utilised in the fight against the spread of coronavirus pandemic in Anambra state; and discover the impact of social media in the fight against the spread of coronavirus pandemic in Anambra State. It was anchored on Agenda Setting Theory, and the Technological Determinism Theory (TDT). The study was designed as a survey with close-ended questionnaire distributed to 400 respondents. The findings of this study revealed that usage and accessibility of social media increased in Anambra state because of coronavirus pandemic. It also revealed that the social media is being utilised by individuals, NGOs and government in the fight against the spread of coronavirus in Anambra state. The study also found that the social media is being utilised to gather and disseminate information, study, transact businesses, among other things. The finding also showed that the social media has positive impact in the fight against the spread of coronavirus in Anambra state. The study concluded that social media has much benefits than negative impact, and should be used to contain the spread of coronavirus. It, among other things, recommended training and empowerment of the citizens on effective utilisation of social media to create impact in the society.

**Introduction**

The Social Media is obviously dominating other mass media of communication all over the universe. Also referred to as New media or Social networking sites, social media include some websites and applications designed, hosted and powered by the Internet to enable people (the users) interact, transact, share information and ideas. Some of them are Badoo, Bumble,

Buzznet, Classmates, Facebook, Flickr, Friendica, Friendster, Happn, Hotor Not, and Hi5. Others are Instagram, Kik, Kuaishou, Likee, LINE, LinkedIn, Live Journal, Lovoo, Moco Space, Musical.ly, My Space, Ning, Ok Cupid, and Pinterest. There are also Quora, Reddit, Zoosk, Zoom, YouTube, WhatsApp, WeChat, VK, Viber, Twoo, Twitter, Tumblr, Tinder, TikTok, Telegram, Tagged, Snapchat, StumbleUpon, Skype, Skout, and Signal. (Hudson, 2020; Gkgigs, 2021). Its popularity is probably because it performs virtuously all the functions being performed by the traditional or mainstream media (radio, televisions, newspapers and magazines) at low-cost, timely and users' convenient. Among their functions are news dissemination, cultural propagation, status conferral, agenda setting, education and entertainment. White (2021) identifies its other functions to include provision of safety tips, prevention of misinformation, fundraising and collaboration. This is possible because Gkgigs (2021) observes that 60% of the world population are on Social Media.

Experts believed that the social media has impact on the world population. Some of them including Gonzalez-Padilla & Tortolero-Blanco (2020), Ahmed (2020), Taylor (2020) and Okwodu (2020) disclose that the impact was more felt at the wake of coronavirus (COVID-19) pandemic which was first reported at Wuhan, China on December 2019. The pandemic, although still ravaging the world, is reported to have increased the numbers of social media users astronomically. It was on Social Media the World Health Organisation (WHO), various countries National Centre for Disease Control (e.g. known in Nigeria as the Nigeria Centre for Disease Control, NCDC), States governments and others broke the news of the virus, how they could be prevented, and other relevant information. It was also utilized in popularising coronavirus terms or jargons like social distancing, physical distancing, lockdown, vaccines, face or nose masks, Personal Protective Equipment (PPE), ventilator, etc. (White, 2021; Thornton, 2020; Upham, 2020; Dolcourt, 2020).

Most of these organisations and states either designed or redesigned their websites and myriads of social media platforms for timely/daily update of their members/citizens on COVID-19 and other related matters. In Nigeria, specifically, the federal government, the 36 states government and organisations, also employed it to appeal for financial or material supports from the public in order to arm their members well to combat and contain spread of the virus. They also utilised it to broadcast or disseminate information about helps, donations or supports received from individuals, religious institutions, and corporate entities. It is against this background that this study intends to assess the impact of social media in the fight against the spread of coronavirus (COVID-19) pandemic in Anambra state, Nigeria.

**Statement of Problem**
The is no gainsaying the facts that coronavirus pandemic brought some untold hardship or challenges to people all over the world. Some markets, offices, private and public institutions were temporary shutdown especially during the lockdown; thereby rendering many persons jobless. This resulted to increase in domestic violence, hunger, starvation, robbery, rape, unplanned pregnancies, as well as talents and technological discoveries. Social media, a technological hub, was among the media of communication utilised in providing people with detailed information about COVID-19 and their effects on people.

Social media provided both factual and fake information, promoted new business opportunities and transactions, services and ideas. These impacted either positively or negatively on the fight against the spread of coronavirus. Hence, this study aims to unravel the impact of social media in the fight against the spread of coronavirus in Anambra state, Nigeria.

**Objectives of Study**
This study was guided by the following objectives;
- ➢ To find out if the numbers of social media users increased in Anambra state since the wake of coronavirus pandemic.
- ➢ To find out if the social media is being utilised in the fight against the spread of coronavirus pandemic in Anambra state.
- ➢ To find out how the social media is being utilised in the fight against the spread of coronavirus pandemic in Anambra state.
- ➢ To find out the impact of social media in the fight against the spread of coronavirus pandemic in Anambra State.

**Research Questions**
This study shall consider the following questions converted from the above research objectives;
- ➢ Are there increase in the numbers of social media users in Anambra State since coronavirus pandemic started?
- ➢ Is social media utilised in the fight against the spread of coronavirus pandemic in Anambra State?
- ➢ How is social media utilised in the fight against the spread of coronavirus pandemic in Anamnbra state?
- ➢ What are the impact of social media in the fight against the spread of coronavirus pandemic in Anambra state?

**Literature Review**
The popularity of Social Media in information dissemination is no longer in doubt. It is not only known as mass medium but also a technological hub where series of activities, ideas, businesses, interactions and actions are born, nurtured, built on, executed and reaped (utilised). This could be what Marshall McLuhan, in 1964, described as a global village. In Nigeria alone, there are 30.95 million active social media users (Tankovska, 2021). With social media, people of all ages, races, religious groups, languages and cultures, educated and uneducated, literates, semiliterate and illiterates dine and wine together. Hence, Osadolor (2019) argues that it introduced 'a new economy' and overcomes all forms of restrictions to press freedom. This new economy is surely a business empire housing various opportunities or golden opportunities.

This is why Togun (2020) states that the social media has impact on the society. She listed some of these impacts as public accountability, channels for citizen participation in decision making and governance, and advocacy. She cited a case of a viral video that showcased how a Nigerian politician wrongly disposed a face mask he wore while attending a burial of top federal government official suspected to have died of COVID-19 complications last year. The reactions of social media users attracted government to redispose the face mask and sanitise the location. Togun also cited how some viral pictures of social events organized by a movie star and artists made the federal and a state government to prosecute them last year over violation of the COVID-19 protocols.

Social media is also a place where everyone is an expert, journalist, author, publisher, business executive, professor, etc. Thus, both accurate and inaccurate information on how to observe coronavirus safety protocols, numbers of deaths recorded due to COVID-19 complications, vaccines, and others were disseminated online by all kinds of persons, news sites, etc. (Togun, 2020; White, 2021; Ahmad and Murad, 2020).

Some of the information on social media related to coronavirus, elicit fear, panic, anxiety and depression (Brewer, 2020; Rothschild, 2020). On January 26, 2021, specifically, the police paraded a politician (social media user) for allegedly spreading a viral message which accused some top government officials in Anambra state, Nigeria of contracting and spreading the virus. Onuegbu (2021) quoted the command's public relations officer, Mr Haruna Muhammed of accusing the suspect of originating and publishing malicious publications to incite and breach public peace in the state.

Again, social media is also useful in carrying out collaborative actions. The Anambra state government, for instance, used it for mobilising funds and other donations aimed at procuring materials for containing the spread of the coronavirus. The names of donors, items donated and other information were constantly updated on the websites of the Anambra Broadcasting Service (ABS) and the state government websites respectively. It was also observed that the links containing the data were being shared across other social media platforms by government agents and other interested persons.

Again, some nongovernmental organisations (NGOs), individuals and civil society organisations (CSOs) in Anambra state used it to mobilize funds and other logistics they shared to the populace during the COVID-19 lockdown. Among them are the Anambra State COVID-19 Civil Society Network, COVID-19 Citizens Support groups, Let's Build Ebenebe (LBE) Forum, and Umuaba Youth Ebenebe, founded on WhatsApp. The able to do members of these CSOs and NGOs operating on WhatsApp, contributed money, bought face masks, hand sanitisers, some cartons of noodles and 10 KG bags of rice and shared to some less priviledged members of the society during and after the lockdown. Some of them also embarked on house-to-house campaigns to sensitise people on how to observe COVID-19 safety protocols.

Similarly, the social media changed the venues and the strategies for conducting teaching and learning, religious and social gatherings in order not to break the rule of social distancing and other safety protocols, especially during the coronavirus lockdown. Conferences, seminars, workshops, lectures, church services, ceremonies like birthday, and others were held using video and audio enabled interactive platforms like Zoom, Google Meet, Teams, Skype, Facebook Room, Slide, YouTube, etc. (Bere, 2020; Orjinmo, 2020; The Conversation, 2020; Okocha, 2020; König, Jäger-Biela & Glutsch, 2020; Shahzad, Hassan, Aremu, Hussain and Lodhi, 2020). Some states government including Anambra, Lagos, and Enugu, respectively launched lessons for primary and secondary schools' students using their state owned radio and television stations (and their websites and social media handles).

The social media has also proved to be an effective tool for initiating an action. There is possibility that such an action would become a reality or come to pass in the physical world. This entails why Nigeria was literary shutdown from the month of October to November 2020 owing to a nationwide mass protest tagged #EndSARS. The hashtag, #EndSARS was a campaign against a police unit, Special Anti-Robbery Squad (SARS), and was actually started in 2016 by one Segun Awosanya (alias Segalink). Segalink started the hashtag on Twitter, a social media platform, before being replicated on other platforms such as Facebook, Instagram, among others (Iwenwanne, 2018).

As it degenerated to street protests, social media was employed to mobilise people to stage (organise) or join the protests at locations close to them, solicit and contribute funds for it successful executions. Some went further to launch an online radio station, Soro Soke (a Yoruba phrase for 'Speak Up') which they used for live broadcast of their activities (Bankole, 2020; BBC News, 2020).The Federal Government eventually disbanded SARS, formed Special Weapons and Tactics (SWAT), as well as ordered for a panel of inquiry to be setup across the 36 states of the federation and Federal Capital Territory (Abuja) to entertain the cases of brutality and extrajudicial killings allegedly perpetuated by the defunct SARS officers, and proffer solutions on how best the victims should be compensated (Akinwotu, 2020; Prince will, 2021).

Perhaps, it was also utilized by 'hoodlums' accused of hijacking the protest in moblising their members to burn down police stations, killing of security officers and law enforcers, vandalising, as well as robbing financial institutions, and others. It was also observed that during the mayhem that followed the protest, users of social media utilised it in mobilising people to where unshared palliatives meant for distribution during the coronavirus lockdown were kept so they could easily be busted and shared.

**Empirical Review**

This study reviews five research works related to the impact of social media in the fight against the spread of coronavirus (COVID-19) pandemic in the world. Adichie (2021) examined the impact of COVID-19 pandemic on the Roman Catholic Church in South Eastern (region), Nigeria, using qualitative research method. The research was conducted between August 29 to September 27, 2020. 18 respondents from 12 Dioceses within the five South Eastern states namely, Anambra, Abia, Enugu, Imo and Ebonyi were interviewed over the phone through audio recording and WhatsApp. The data were collected using the Key Informant Interview (KII). It was found, among other things, that the pandemic led to alterations of some liturgical celebrations, forceful compliance to safety protocols, financial constraints, and backsliding.

Obi-Ani, Anikwenze, and Isiani (2020) conducted an investigation entitled, the social media and the Covid-19 pandemic: observations from Nigeria. The methodology for the research design was field survey which enabled the researchers to collect data from primary and secondary sources. The data were generated through personal communication, telephone conversations, online personal messages, newspapers, journal articles and online materials. A total of 100 questionnaires were equally distributed to respondents on different online platforms and personal contacts. It was found that the social media has both positive and negative impact in preventing coronavirus in Nigeria.

BearingPoint (2020) conducted an in-depth study to find out Africa's Digital Solution to tackle COVID-19. The study was conducted from May to June, 2020 under the supervision of the European Investment Bank, and with collaboration of the United Nations Development Programme. 14 participants from different countries in Africa assisted in the study. It employed interviews and questionnaires as means of data collection. Data was gathered from 31 out of 54 African countries such as Angola, Benin, Central Africa Republic, Chad, Comoros, and Republic of Congo. Others were Djibouti, Equatorial Guinea, Eswatini, Ethiopia, Gambia, Guinea, Kenya, Lesotho, Liberia, Mali, Mauritania, and Mauritius. Also included were Morocco, Mozambique, Niger, Nigeria, Rwanda, Senegal, Sierra Leone, South Africa, South Sudan, Togo, Tunisia, Uganda and Zimbabwe. It was found that many African countries devised and utilised digital solutions including the social media in monitoring and combating the spread of coronavirus pandemic.In other words, the social media impacted positively in combating the spread of coronavirus pandemic in Africa.

Ahmad and Murad (2020) investigated the impact of social media on panic during the COVID-19 pandemic in Iraqi Kurdistan using online questionnaire (quantitative methodology). Objective of the study was to ascertain how social media affects mental health and the spread of panic about COVID-19 in the Kurdistan Region of Iraq. 516 social media users attempted the questionnaire, and the data was analysed using the SPSS software. It was found that the social media has a significant impact on spreading fear and panic related to the COVID-19 outbreak, as well as influence people's mental health and psychological well-being negatively.

Dhanashree, Chauhan, Bhatia, Sethi and Chauhan (2020) examined the role of mass media and its impact on general public during coronavirus disease 2019 pandemic in North India: An online assessment. The study adopted an online survey using a questionnaire drafted on Google spreadsheets. The questionnaire was sent (circulated) to the researchers' contacts (respondents between the ages of 10 and above) in North Indian states between June 23 to July 3, 2020. A total of 384 responses were gathered at the end. It was found that the social media was not only the most used or accessible mass medium in India but also create anxiety, fear, panic, among other impact on users.

**Theoretical framework**
The theoretical underpin for this study are Agenda Setting Theory and Technological Determinism Theory (TDT). The two are used in order to bridge the gap or cover any lapses either of them could have while exploring the impact of social media in the fight against the spread of coronavirus (COVID-19) pandemic in Anambra state, Nigeria.

**The Agenda Setting theory**
The Agenda Setting theory believes that the media sets agenda for the society or people by promoting any issue of importance. The issue could be on health, politics, economy, rural or urban life or people. By making such an issue popular, the media influences the audience to think or behave in a certain way it (the media) so desired.

Developed in 1968 by Maxwell McCombs and Donald Shaw, the theory also argues that most issues discussed in the society are those already discussed, projected or popularised by the media (Blood, 1982; McCombs & Reynolds, 2002; Wogu, 2008). That is to say that once the media starts giving favourable attention or prominence to an issue, the public would no doubt que in and make it a public discourse.

Communication (2018) identifies two assumptions of agenda setting theory as to filter and shapes societal views, and to make society to consider things as important. Some of the ways the media set agenda for society are through phone-in or audience participatory programmes, drama, feature stories, documentaries, news, among others. In social media, specifically, agenda could be set by creating a hashtag for the issue or reposting (retweeting) it on other platforms and accounts (Demirsoy and Karakoc, 2016). Hence, Demirsoy and Karakoc insist that increase in numbers of social media users is evident that it sets agenda to society.

This was observed immediately the WHO announced the outbreak of coronavirus at China. There was a reported increase in numbers of social media users all over the world, including in Anambra state, Nigeria. These new and old users of social media took to cyberspace to learn the safety protocols, discover the names and pictures of some victims, identifiable signs or symptoms of the pandemic and other jargons related to it as popularised by the social media.

**Technological Determinism Theory (TDT)**

Technological Determinism Theory is believed to have been coined by an American's Social Scientist, Thorstein Veblen (1857–1929). It is anchored on the believes that the new or emerging technology shape the society. It also believes that although technology is developed by humans, it in turns, influences and shapes the human feelings, behaviour or attitude (Anunike & Onuegbu, 2020; Culkin, 1967).

Little wonder, Sasvari (2012) states that "the information and telecommunication technologies play a constantly expanding role in all fields of social existence, which has shaken the foundations of social structures and processes and resulted in profound changes in politics, economy, culture, and everyday life." Social media is one such information and telecommunication technology developed by humans. It is also influencing or impacting the creators (the world/the humans) positively and negatively. In healthcare, it's used for data or information gathering from the patient, observation, self-assessment, checkup and cure (Gücina and Özlem, 2015).

**Methodology**

**Research Design**
The research design adopted by the researcher was survey. This research design was apt for the study due to the research objectives.

**Area of Study**
The area of study was Anambra State. Anambra State is among the five States in the South East geopolitical zone of Nigeria. The State consists of 181 communities, politically, grouped into 326 electoral wards, 21 Local Government Areas (LGA) and 3 Senatorial Districts (zones). Awka is the State capital, while Onitsha, Nnewi, Ekwulobia, Nkpor, and Ogidi are some of her prominent towns.

**Population of the Study**
The population of the study was all the Anambra State residents. The state population is 10.8million people as at 2019 (Anambra, 2021).

**Sample Size and Sampling Procedure**
The sample size for the study was 400. This was based on Taro Yamane's formula for sample size determination:

$$n = \frac{N}{(1 + N[e]^2)}$$

Where n = Sample size
N = Population
e = Error margin

That is;  $\dfrac{10,800,000}{(1+10,800,000\ [0.05]^2)}$

$= \dfrac{10,800,000}{1+ (10,800,000 \times 0.0025)}$

$= 1+ 27,000$

$= 27,001$

$= \dfrac{10,800,000}{27,001}$

$= 399$

Therefore, the sample size is 399 at 5 percent error margin. But it is further approximated to 400 to achieve mathematical convenience.

The study also adopted simple random probability sampling. It is appropriate for the study because it affords every member of the population equal chance to participate in the survey.

To achieve this, the researcher listed the names of 3 Senatorial zones in the state and their LGAs as follow:
1. Anambra Central has Anaocha, Awka North, Awka South, Dunukofia, Idemili North, Idemili South, and Njikoka local government areas;
2. Anambra North contains Anambra East, Anambra West, Ayamelum, Ogbaru, Onitsha North, Onitsha South, and Oyi local government areas;
3. Anambra South has Aguata, Ihiala, Nnewi North, Nnewi South, Orumba North, and Orumba South local government areas.

The researcher randomly selected two LGAs from each of the senatorial zones. They are Anambra Central (Awka South and Njikoka LGAs); Anambra North (Onitsha North and Oyi LGAs)and Anambra South (Aguata and Orumba South). That is, six Local Government Areas were selected. A visit was paid to each of the Local government headquarters namely, Amawbia (Awka South), Abagana (Njikoka), Onitsha (Onitsha North), Nteje (Oyi), Ekwulobia (Aguata) andUmunze (Orumba South). 66copies of questionnaire were administered on respondents found in each of the Council Area (headquarters), totaling 396. To make it 400;

the sample size, the remaining four copies of the questionnaire were administered on respondents at Amawbia, the headquarters of Awka South LGA. This is because the Council Area is the State capital city.

**Instrument of Data Collection**

The study used questionnaire to collect data from respondents. The questionnaire has two sections: section one seeks the respondents' personal data; while section two seeks to find out the impact of social media in the fight against the spread of coronavirus.

### 3.1.6 Pre-test of Instrument

To test the validity and reliability of the instrument for this survey, the researcher conducted a pilot study using 30 respondents randomly selected from the population. The respondents filled and returned the questionnaire. Analysis of their answers convinced the researcher of his instrument validity.

Ten days later, the researcher re-administered the same instrument to the same 30 respondents to confirm its reliability (consistency). It was also analysed. The analysis showed that there was no significant different between them and the answers they supplied in the first exercise. This convinced the researcher of his instrument reliability.

**Method of Data Analysis**

The method of data analysis used in this study was frequency tables and simple percentages.

**Data Presentation and Analysis**

A total of 400 copies of the questionnaire was administered. But 321, representing 80% were returned, while 79 representing 20% were unrecovered. The results were presented as follow:

**Section A:**

Table I: Respondents Demographic Characteristics

| **Responses/Categories** | **Frequency** | **Percentages (%)** |
|---|---|---|
| Gender | | |
| Male | 148 | 46% |
| Female | 173 | 54% |
| **TOTAL** | **321** | **100%** |
| Age Group | | |
| 18-35 | 107 | 33% |
| 36-50 | 155 | 48% |
| 51-above | 59 | 18% |
| **TOTAL** | **321** | **100%** |
| Education Background | | |
| First School Leaving Certificate (FSLC) | 42 | 13% |
| National Certificate on Education (NCE)/Ordinary National Diploma (OND) | 99 | 31% |
| Bachelor Degree (B. Sc./B.A/B.Ed/LL.B/HND) | 116 | 36% |
| Postgraduate Degree/Diploma (PDE/PGD/M.Sc./M.A/ LL.M/PhD.) | 64 | 20% |
| **TOTAL** | **321** | **100%** |
| Marital Status | | |
| Single | 165 | 51% |
| Married | 81 | 25% |
| Divorced | 5 | 2% |
| Widow/Widower | 70 | 22% |
| **TOTAL** | 321 | 100% |

Table 1 identifies the respondents' demographic characteristics. It reveals that 54% (n=173) of them are females, while 46% (n=148) are females. Females population are more than males population because they seemed to have more interest on the impact of social media.

The table further shows that most of the respondents are between the ages of 36-50 which has a total number of 55 representing 48%. Others are respondents aged 18-35 which constitutes 33% (n=107), and those aged 51 and above are 18% (n=59).

The respondents education attainment indicates that 42 of them constituting 13% are First School Leaving Certificate (FSLC) holders. Also, 31% (n=99) have acquired National Certificate on Education (NCE) and Ordinary National Diploma; 36% (n=116) are first degree holders in different fields; while the postgraduate degree holders among them are 20% (n=64).

The table also presents the respondents marital status. It shows that 51% (n=165) of them are unmarried (single), 25% (=81) are married, just as those divorced are 2% (n=5), while the widows/widowers are 22% (n=70).

**Section B:**

Table 2: Social media users

| Question: | Responses | Frequency | Percentages (%) |
|---|---|---|---|
| Do you have an account (profile )on any of the social media platforms? | Yes | 321 | 100% |
| | No | 0 | 0% |
| | **Total** | **321** | **100%** |

Table 2 shows that all the respondents that participated in the study have knowledge of the social media. The 321 respondents representing 100% admit that they are on social media.

Table 3: Mostly accessed social media platforms in Anambra state

| Question: | Responses | Frequency | Percentages (%) |
|---|---|---|---|
| If 'yes,' name the social media platforms you access mostly | Facebook | 79 | 25% |
| | Instagram | 31 | 10% |
| | Twitter | 56 | 17% |
| | YouTube | 67 | 21% |
| | WhatsApp | 88 | 27% |
| | **Total** | **321** | **100%** |

Table 3 identifies the social media platforms accessed by the respondents. 25% (n=79) of them are on Facebook, 10% (31) are on Instagram, while the Twitter users are 17% (n=56). Others are users of YouTube constituting 21% (n=67), and 27% (n=88) WhatsApp users.

Table 4: Accessibility of social media platforms before COVID-19

| Question: | Responses | Frequency | Percentages (%) |
|---|---|---|---|
| How often do you access your social media accounts (profile) before the COVID-19? | Sometimes | 74 | 23% |
| | Often | 144 | 45% |
| | Very often | 103 | 32% |
| | **Total** | **321** | **100%** |

Table 4 shows the respondents level of exposure or accessibility of the social media before the COVID-19 pandemic. 23% (n=74) of them do that sometimes, probably when they are less busy or opportune. 45% (n=144) of the respondents access their social media accounts often, while 32% (n=103) do that always (very often).

Table 5: Accessibility of social media platforms since COVID-19 started

| Question: | Responses | Frequency | Percentages (%) |
|---|---|---|---|
| How often do you access your social media accounts (profile) since COVID-19 started? | Sometimes | 13 | 4% |
| | Often | 81 | 25% |
| | Very Often | 227 | 71% |
| | **Total** | **321** | **100%** |

Table 5 proves that more respondents are accessing their social media accounts since COVID-19 pandemic. 4% (n=13) of them don't always (sometimes) access their social media accounts or profiles, 25% (n=81) do that often, while 71% (n=227) respondents do that very often.

Table 6: Reasons for change in usage of social media during COVID-19

| Question | Reasons |
|---|---|
| State why your presence on social media seemed to have increased or reduced (declined) during COVID-19 pandemic | Academics/Events |
| | Financial constraints |
| | Business transactions |
| | Information gathering/dissemination |
| | Fake news/Misinformation/Disinformation |
| | Loneliness/Interactions |

Table 6 summarises the reasons given by respondents on why they either increased or reduced the number of time they access their social media accounts (profiles) during COVID-19 pandemic. It is an open-ended question. They include academics/events, financial constraints, business transactions, information gathering and dissemination, fake news, misinformation, disinformation, loneliness and interactions.

Table 7: Social media and spread of COVID-19

| Question: | Responses | Frequency | Percentages (%) |
|---|---|---|---|
| Do you think the social media is being utilised by the government, NGOs or individuals in the fight against the spread of COVID-19 pandemic in Anambra state? | Yes | 289 | 90% |
| | May be | 32 | 10% |
| | No | 0 | 0% |

|  | Total | 321 | 100% |
|---|---|---|---|

Table 7 shows that most of the respondents agree that the social media was utilised by individuals, nongovernmental organisations (NGOs) and the government in the fight against the spread of coronavirus (COVID-19) pandemic in Anambra state. 90% (n=289) of them answered yes to the questionnaire, 10% (n=32) chose 'no', while none tick the 'no' option.

Table 8: Utilisation of social media in the fight against the spread of COVID-19 in Anambra state

| Question: | Responses | Frequency | Percentages (%) |
|---|---|---|---|
| If your response was 'yes' or 'may be,' why? | Information Gathering/Dissemination | 98 | 31% |
|  | Collaborations | 30 | 9% |
|  | Funds/Logistics Mobilisations | 78 | 24% |
|  | Education | 96 | 30% |
|  | Events/Sensitisations | 19 | 6% |
|  | Total | 321 | 100% |

Table 8 presents respondents' views on why they believe the social media is being utilised in the fight against the spread of coronavirus pandemic in Anambra state. 31% (n=98) say the social media was used for information gathering and dissemination, while 9% (n=30) maintain that it is used for seeking collaborations. 24% (78) agree it is used for funds/logistics mobilisations, 30% (n=96) are for education, and 6% (n=19) are for events.

Table 9: General impact of social media in the fight against spread of COVID-19

| Question: | Responses | Frequency | Percentages (%) |
|---|---|---|---|
| What do you consider as the impact of social media in the fight against the spread of coronavirus pandemic in Anambra State, Nigeria? | Positive | 178 | 55% |
|  | Negative | 143 | 45% |
|  | Total | 321 | 100% |

Table 9 shows that social media has positive impact in the fight against the spread of coronavirus in Anambra state. 55% (n=178) of the respondents agree that it has positive impact, while 45% (143) say it has negative impact.

**Answering the research questions**

**Research Question One:**

Are there increase in the numbers of social media users in Anambra State since the coronavirus pandemic started?
Data in table 5 showed that usage and accessibility of social media increased in Anambra State since the COVID-19 pandemic. This was a bit higher compared to the data in table 4. Table 6 also addressed the question as respondents gave reasons for their actions to include online teaching and learning, and participation on online seminars and workshops (academics/events), business transactions, information gathering and dissemination, and others.

**Research Question Two:**
Is social media utilised in the fight against the spread of coronavirus pandemic in Anambra State?
Table 7 proved that the social media was utilised in the fight against the spread of coronavirus pandemic in Anambra state. 90% (n=289) of the respondents responded affirmatively that individuals, nongovernmental organisations (NGOs) and the government utilised the social media in the fight against the spread of coronavirus (COVID-19) pandemic in Anambra state.

**Research Question Three:**
How is social media utilised in the fight against the spread of coronavirus pandemic in Anambra state?
Table 8 answered the research question. The data in the table showed that the social media is being utilised in the fight against the spread of COVID-19 in the following ways; gathering and dissemination of COVID-19 related information, executing collaborative actions on COVID-19, mobilisations of funds/logistics, education including academic purposes, and events.

**Research Question Four:**
What are the impact of social media in the fight against the spread of coronavirus pandemic in Anambra state?
Table 9 answered the research question. It shows that social media has both positive and negative impact. However, the positive impact is more as 55% respondents agreed that it was impacting positively in the fight against the spread of coronavirus in Anambra state.

**Conclusion**
It was found at the course of this study that the social media gained much membership and accessibility in Anambra state since emergence of the coronavirus pandemic. The finding also proved that the social media is being utilised to execute several COVID-19 related activities including information gathering, events, collaborations, education and others. These findings are related to those found by Obi-Ani, Anikwenze, and Isiani (2020), BearingPoint (2020), and Okocha, 2020. The study hereby concludes that the social has much positive impact than its negative impact in the fight against the spread of coronavirus (COVID-19) pandemic in Anambra state.

**Recommendations**

The following recommendations are necessary based on this research findings:
1. Nigerians, especially those resident in Anambra State, should strive to equip themselves academically and otherwise so as to effectively fit in, in the cyberspace where the social media is domicile. Communication, computer application, graphics arts and design, and other skills are helpful in using the social media to impact the society.
2. Social media users should also seek, study and understand relevant laws of the land, including the Freedom of Information Act, 1999 constitution of the federal republic of Nigeria, Evidence Act, and others, and apply them accordingly while accessing the social media. This will help to prevent misinformation, fake news, character defamations, among others.
3. There is also need for social media users to have basic knowledge of mass communication or journalism. There are so many online academic institutions, books, journals, news reports, tools and others capable of equipping them to become good communicators, and to communicate effectively without creating panics, fear and anxiety.
4. Similarly, there is need for the states and the federal government of Nigeria, and politicians to embark on human capital development and empowerment of the citizens through the social media. They should note that no one armed with the knowledge of social media would ever be hungry.
5. Health practitioners, NGOs and CSOs in Anambra state should create functional websites, blogs and social media handles, and update their activities therein. This is also applied to the federal and states government ministries, departments, agencies and parastatals. They should also employ the services of practicing journalists for wider dissemination of their information through the mass media. This will help both researchers and non-researchers to gain quick access to relevant information from them when the need arises.

Webology (ISSN: 1735-188X)
Volume 19, Number 2, 2022Akinwotu, E. (2020, October 11). Nigeria to disband SARS police unit accused of killings and brutality. https://www.theguardian.com/world/2020/oct/11/nigeria-to-disband-sars-police-unit-accused-of-killings-and-brutality

Anambra, S. (2021). Anambra state population. https://anambrastate.gov.ng/

Anunike, O. & Onuegbu, O. (2020). Youth's exposure and utilization of Internet advertisements in Awka, Anambra State. Journal of Communication and Media Studies, 1 (2), 254-269.

Bankole, I. (2020, October 17). Listen: #EndSARS launches online Radio 'Soro Soke' (audio). https://www.vanguardngr.com/2020/10/listen-endsars-launches-online-radio-soro-soke-audio/

BearingPoint (2020, July). Africa's Digital Solution to tackle COVID-19. Luxembourg: European Investment Bank. https://www.eib.org/attachments/country/africa_s_digital_solutions_to_tackle_covid_19_en.pdf

Bere, G. (2020, March 30). COVID-19: Christians comply with directive, hold service at home.https://www.sunnewsonline.com/covid-19-christians-comply-with-directive-hold-service-at-home/

Blood, W. (1982). Agenda Setting: A Review of the Theory. https://doi.org/10.1177/1329878202600101

Brewer, K. (2020, March, 16). Coronavirus: how to protect your mental health. https://www.bbc.com/news/health-51873799

Communication (2018, 19). The Agenda-Setting Theory in Mass Communication. https://online.alvernia.edu/articles/agenda-setting-theory/

Demirsoy, A. & Karakoç, E. (2016, April 4). Contribution of social media to agenda setting approach. https://dergipark.org.tr/tr/download/article-file/397974

Dhanashree, G. H., Chauhan, A., Bhatia, M., Sethi, G., Chauhan, G. (2020). Role of mass media and it's impact on general public during coronavirus disease 2019 pandemic in North India: An online assessment. Indian J Med Sci. doi: 10.25259/IJMS_312_2020

Dolcourt, J. (2020, April 8). Coronavirus glossary: every COVID-19 related term you need to know. https://www.cnet.com/how-to/coronavirus-glossary-every-covid-19-related-term-you-need-to-know/

Gkgigs (2021, March 2). List of all social networking sites and their founders. https://www.gkgigs.com/list-of-all-social-networking-sites-and-their-founders/

Gonzalez-Padilla, D. A. & Tortolero-Blanco, L. (2020). Social media influence in the COVID-19 pandemic. Spain International braz j urol, 46(1). https://doi.org/10.1590/s1677-5538.ibju.2020.s121

Gücina, N. O. &Özlem, S. B. (2015). Technology Acceptance in Health Care: An Integrative Review of Predictive Factors and Intervention Programs. Procedia- Social and Behavioral Sciences, 195, 1698–1704.

Hudson, M. (2020, June 23). What is Social Media: Definitions and Examples of Social Media. https://www.thebalancesmb.com/what-is-social-media-2890301

Iwenwanne, V. (2018, November 13). Civic Tech startups take on Nigeria's struggle for justice. https://www.devex.com/news/civic-tech-startups-take-on-nigeria-s-struggle-for-justice-937516385                                          http://www.webology.org

**Appendix**
**Questionnaire**

Please, tick appropriately (√)

**Section A: Personal data**

Gender: A. Male ( )     B. Female ( )

Age: A. 18-35 ( )     B. 36-50 ( )     C. 51-above ( )

Education background: A. FSLC ( )   B. SSCE ( )   C. NCE/ND ( )     D. HND/B. Sc. ( )     E. PG ( )

Marital Status: A. Single ( )     B. Married ( )     C. Divorced ( )     D. Widow/Widower ( )

**Section B**

1. Do you have an account on any of the Social Media platform? A. Yes ( )   B. No ( )
2. If 'yes,' name the social media platforms you have access to…………………………
3. How often do you access your social media account (s) before the COVID-19 pandemic?   A. Sometimes ( )     B. Often ( )     C. Very Often ( )
4. How often do you access your social media account(s) since the COVID-19 pandemic? A. Sometimes ( )     B. Often ( )     C. Very often ( )

5. Do you think the social is being utilised by the government, NGOs or individuals in the fight against the spread of COVID-19 pandemic in Anambra state? A. Yes ( ) B. No ( )
6. If 'yes,' how is it utilised in the fight against the spread of COVID-19 in Anambra state? A. Information dissemination ( )   B. Education ( )   C. Collaboration ( )
7. Generally, what do you consider as the impact of social media in the fight against the spread of coronavirus pandemic in Anambra State, Nigeria? A. Negative   ( )   B. Positive ( )